# Semantic Document Clustering on Named Entity Features

Tru H. Cao, Vuong M. Ngo, Dung T. Hong, and Tho T. Quan

*Abstract*—**Keyword-based information processing has limitations due to simple treatment of words. In this paper, we introduce named entities as objectives into document clustering, which are the key elements defining document semantics and in many cases are of user concerns. First, the traditional keyword-based vector space model is adapted with vectors defined over spaces of entity names, types, name-type pairs, and identifiers, instead of keywords. Then, hierarchical document clustering can be performed using the similarity measure defined as the cosines of the vectors representing documents. Experimental results are presented and discussed. Clustering documents by information of named entities could be useful for managing web-based learning materials with respect to related objects.**

## I. INTRODUCTION

As witnessed, the World Wide Web has become a huge and important knowledge treasure of humankind, bringing important changes in many areas, including education. In order to exploit that on-line electronic resource, tools for searching and managing web documents are indispensable. Much research has recently embarked on the issue of integrating data mining techniques and pedagogical theories. Studies of web usage analysis have been applied to e-learning ([15]). Some research systems and prototypes like CourseVis ([9]) and TADA-Ed ([10]) offer visualization tools allowing teachers to explore student data. SSWeb ([12]) is an attempt that makes use of clustering techniques and the Semantic Web technologies to provide scholarly retrieval services for those who are seeking scholarly materials in research and study.

In particular, for vast information resources currently available on the Web, clustering methods, combined with information retrieval techniques, are deemed highly useful for learners to observe and infer knowledge. For example, a comprehensible presentation of clustered Web pages about cities in the world can furnish observers with political, economical, historical and geographical information in an intensive and impressive manner. From this, one can efficiently acquire knowledge about business centers in a certain region like East Asia, for instance. Current search engines such as Google are useful for finding documents containing certain keywords. Furthermore, due to the deficiencies of the query-list approach to showing search results, research has been carried out, and systems implemented, to group returned documents into meaningful thematic categories ([16], [11]).

However, since key words or phrases processed by simple string matching are not adequate to deal with the semantics of documents, those raw text-based searching and clustering suffer disadvantages. For example, if we want to find documents about cities by posing the keyword "*city*", Google will not return those documents that do not contain that keyword, though they include true city names like "*Shenyang*" or "*Beijing*". On the other hand, when we want to find documents about *Shenyang as a city* it may wrongly return also documents about *Shenyang Red River* or *Shenyang Qipanshan Mountain*, just because they contain the keyword "*Shenyang*". As another fact, keyword-based clustering techniques cannot separate geographical documents about rivers and those about mountains, for instance.

That is actually the common disadvantage of the current Web generation, which is for human consumption rather then machine processing. Therefore, the Semantic Web has emerged to embed semantics into raw text web pages, so as to make them machine-understandable and facilitate more intelligent information processing than with the current Web technologies ([2]). In particular, the above-mentioned shortcomings of keyword-based document searching and clustering can be overcome if named entities (NE) occurring in documents are marked up with their types, i.e. river or mountain, for instance. In fact, automatic NE recognition is one of the basic issues for the Semantic Web and has attracted much research effort. Semtag ([4]), using statistical methods, and KIM ([8]), using pattern matching rules, are well-known systems to be named.

For semantic enhancement of searching, [3] and [5], among other works, adapted the traditional keyword-based vector space model (VSM) to take into account named entities mentioned in documents. Named entities mentioned in a document are also the key elements defining the document semantics and are what users may be concerned with. As a motivating example, one may want to have Web-based learning materials in geography separated into those about rivers and those about mountains. Furthermore, based on entity names or identifiers, the documents in the river group, can be then clustered into subgroups for each particular river.

In [14], the most significant entity name in a document was used as its label, based on an enhanced version of the *tf.idf* measure. Then the documents with labeling named entities of the same type were grouped together. As such, it was simply classification of documents by the types of their representative entity names, rather than clustering. Consequently, it could not produce a partition each cluster of which was a group of documents having close semantics regarding various named entities occurring in them.

Tru H. Cao, Vuong M. Ngo, Dung T. Hong, and Tho T. Quan are with the Faculty of Computer Science and Engineering, Ho Chi Minh City University of Technology, Vietnam (corresponding email: tru@cse.hcmut.edu.vn).

Therefore, in this paper, we introduce named entities into document clustering, which to our knowledge has not been adequately investigated. In the scope of this paper, as the first step, we are not concerned with the best algorithm for NE-based document clustering. Instead, we adapt the traditional keyword-based VSM to represent a document with respect to different features of the named entities occurring in the document, namely their types, names, and identifiers. Then we apply *k*-means, a popular clustering algorithm, to demonstrate the advantages of document clustering using NE information.

The paper is organized as follows. Section II introduces our general framework of NE-based information processing, in which the keyword-based VSM is adapted with vectors over spaces of entity names, types, name-type pairs, and identifiers. Sections III and IV present the proposed method and its experiments of NE-based document clustering. Finally, Section V concludes the paper and outlines further research.

## II. A Framework of NE-Based Information Processing

We recall that named entities represent individuals that can be referred to by names, such as people, organizations, and locations ([13]). For example, consider the following passage introducing the geography of *Shenyang*:

> "*Shenyang is located in the first of three Northeast China provinces, in the center of Liaoning province. It is situated in the inland area of the Liaodong peninsula. Shenyang is north of the Bohai Sea and southwest of the Changbai mountains.*"

Here, *Shenyang*, *Northeast China*, *Liaoning*, *Liaodong*, *Bohai*, and *Changbai* are named entities.

Each named entity can be recognized and annotated with its appearing name, type and, if existing in a knowledge base (KB) of discourse, identifier. That is, a fully recognized named entity has three features, namely, name, type, and identifier. For instance, a full annotation of *Liaoning* may be the triple ("*Liaoning*", *Province*, *#Province_123*), where *Province* is the entity type and *#Province_123* is its supposed identifier. Due to the ambiguity in a context and performance of a recognition method, a named entity may not be fully annotated or may have multiple annotations. For instance, *Changbai* should be recognized as a mountain in this context, though not existing in the KB and thus having no identifier. Meanwhile, *Shenyang* may be ambiguously recognized as both a city and a university.

In summary, a NE annotation can be in one of the following forms:
1. Only name: when the type is not recognizable.
2. Only name and type: when the identifier is not recognizable.
3. Name, type, and identifier.

We note that the names and types of an entity are derivable from its identifier. Also, based on an ontology and KB, given an annotation of a named entity, there are subsumed ones whose types are super-types of the type in that annotation and whose names are aliases of that entity.

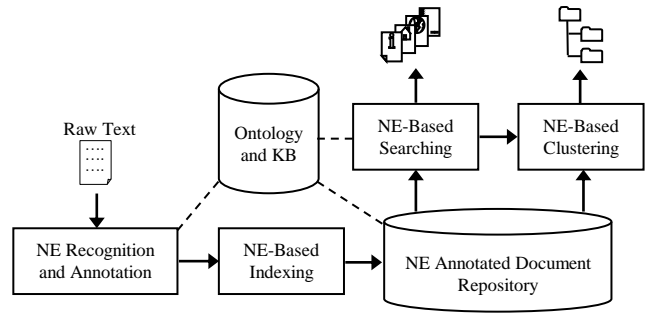

Fig. 1. NE-based information processing

A system of NE-based information processing is shown in Figure 1 It contains an ontology and KB of named entities in a world of discourse. The NE Recognition and Annotation module extracts and embeds information about named entities in a raw text, before it is indexed and stored in the NE Annotated Document Repository. Users can search for documents about named entities of interest via the NE-Based Searching module, or view documents grouped by various features of contained named entities via the NE-Based Clustering module. Documents to be clustered can be from the document repository or the result of a search.

In the keyword-based VSM ([1]), each document is represented by a vector over a space of keywords of discourse. Conventionally, the weight corresponding to a term dimension of the vector is a function of the occurrence frequency of that term in the document, called *tf*, and the inverse occurrence frequency of the term across all the existing documents, called *idf*. As such, VSM is usually associated with the *tf.idf* weighting scheme. The similarity degree between two documents, or a document and a query, can be then defined as the cosine of the two representing vectors.

With terms being keywords, the traditional VSM cannot satisfactorily represent the semantics of documents with respect to the named entities they contain, such as for the following cases:
1. Documents about *cities*.
2. Documents about the *People's Republic of China*.
3. Documents about *Shenyang University*.
4. Documents about the *Shenyang Red River*.

That is because, for Case 1, a target document does not necessarily contain the keyword "*city*", but only some named entities of the type *City*. For Case 2, a target document may mention about the *People's Republic of China* by other names, i.e., the country's aliases, such as "*China*". On the other hand, documents containing entities named "*China*" but different from the country, like China town, are not target documents. For Case 3, documents about *Shenyang* as a city or a hotel are not target documents at all, though containing the keyword "*Shenyang*". Meanwhile, Case 4 targets at documents about a precisely identified named entity, i.e, the Red river in Shenyang, not the one in Hanoi, Vietnam, for instance.

Therefore, we now adapt the traditional VSM with vectors over spaces of entity names (including aliases), types, name-type pairs, and identifiers as follows.

**Assumption II.1** Suppose a triple ($\mathcal{N}$, $\mathcal{T}$, $\mathcal{I}$) where $\mathcal{N}$, $\mathcal{T}$, and $\mathcal{I}$ are respectively sets of names, types, and identifiers of named entities of discourse. Then:
1. Each document (or query) $d$ is modelled as a subset of ($\mathcal{N} \cup \{nil\}) \times (\mathcal{T} \cup \{nil\}) \times (\mathcal{I} \cup \{nil\})$, where *nil* denotes an unspecified name, type, or identifier of a named entity in $d$, and
2. $d$ is represented by the quadruple ($d_\mathcal{N}$, $d_\mathcal{T}$, $d_{\mathcal{NT}}$, $d_I$), where $d_\mathcal{N}$, $d_\mathcal{T}$, $d_{\mathcal{NT}}$, and $d_I$ are respectively vectors over $\mathcal{N}$, $\mathcal{T}$, $\mathcal{N} \times \mathcal{T}$, and $\mathcal{I}$.

Each of the four component vectors introduced above for a document can be defined as a vector in the traditional *tf.idf* model with the only difference in the interpretation of a term being a name, a type, a name-type pair, or an identifier of a named entity, instead of a keyword. That is, let a component vector be $d_\mathcal{K} = (w_{1d}, w_{2d}, \ldots, w_{md})$, where $\mathcal{K}$ and $m$ correspond to $\mathcal{N}$, $\mathcal{T}$, $\mathcal{N} \times \mathcal{T}$, or $\mathcal{I}$, respectively. Then term weights are derived as follows.

Let $N$ be the total number of documents in the system, $n_i$ be the number of documents where a term $k_i$ occurs, and $freq_{id}$ be $k_i$'s raw frequency, i.e., the number of times $k_i$ occurs in $d$. The normalized frequency of $k_i$ in $d$ is defined by:

$tf_{id} = freq_{id} / max_j \{freq_{jd}\}$

where the maximum is computed over all the terms that occur in $d$. If $k_i$ does not occur in $d$, then $tf_{id} = 0$.

Depending on $\mathcal{K}$ being $\mathcal{N}$, $\mathcal{T}$, $\mathcal{N} \times \mathcal{T}$, or $\mathcal{I}$, $k_i$ is said to occur in $d$ if and only if:
1. $\mathcal{K} = \mathcal{N}$, $k_i$ is a name, and $d$ contains a named entity whose appearing name or one of its aliases is $k_i$, or
2. $\mathcal{K} = \mathcal{T}$, $k_i$ is a type, and $d$ contains a named entity whose recognized type is the same as, or a subtype of, $k_i$, or
3. $\mathcal{K} = \mathcal{N} \times \mathcal{T}$, $k_i = (n, t)$, and $d$ contains a named entity whose appearing name or one of its aliases is $n$, and whose recognized type is the same as, or a subtype of, $t$, or
4. $\mathcal{K} = \mathcal{I}$, $k_i$ is an identifier, and $d$ contains a named entity whose identifier is $k_i$.

As in the keyword-based case, the inverse document frequency of $k_i$ is defined by:

$idf_i = log(N / n_i)$

While $tf_{id}$ quantifies the occurrence degree of $k_i$ in a particular document $d$, $idf_i$ measures the significance of the occurrence of $k_i$ in every document; the more the number of documents where $k_i$ occurs is, the less significant the occurrence of $k_i$ is. So the weight of $k_i$ to $d$ is defined by:

$w_{id} = tf_{id} \times idf_i$

In brief, the main idea of the proposed NE-based VSM is to adapt the notion of *terms* being keywords in the traditional VSM to be entity names, types, name-type pairs, or identifiers. Then, as usual, the similarity degree of two documents, or a document and a query, can be defined as the cosine of their representing vectors. We note that the join of $d_\mathcal{N}$ and $d_\mathcal{T}$ cannot replace $d_{\mathcal{NT}}$ because the latter is concerned with entities of certain name-type pairs.

In contrast to [3] and [5], which defined only vectors on NE identifier spaces, here we consider spaces of different NE features. The advantage of splitting document representation into four component vectors is that, given a query, searching and matching need to be performed only for those components that are relevant to that query. On the other hand, hierarchical clustering can be performed on different NE features, producing meaningful and useful organization of documents. We also note that, as for any method relying on named entities, the performance of NE-based document clustering depends on that of NE recognition in a preceding stage. However, for research, the two problems should be separated, and this paper's focus is on the former, assuming all named entities in a document correctly annotated beforehand.

### III. HIERARCHICAL DOCUMENT CLUSTERING BY NAMED ENTITIES

In traditional document clustering techniques, key words or phrases are identified and processed to evaluate the document similarity, which is crucial information to perform the clustering process. We now propose to use named entities and show how they can complement raw text-based clustering to make the results more precise and meaningful. NE-based clustering has the following advantages:
1. Named entities can be treated as special terms in which their meanings and relations are precisely pre-defined in an ontology and KB of discourse, whence they can make clustering results more semantically accurate for certain user needs.
2. In some certain domains, such as news published in media and learning materials, named entities suggest meaningful representations for generated clusters because they capture salient points in document contents.

The basic steps for NE-based hierarchical document clustering are as follows:
1. *Named entity recognition*: It recognizes named entities in the documents to be clustered. Recognition algorithms rely on a KB of named entities as in KIM ([8]).
2. *Vectorization*: It vectorizes documents based on the recognized named entities. Each document is represented by the quadruple of vectors as in Assumption II.1, with the weights evaluated using the *tf.idf* model as presented above.
3. *Multi-objective hierarchical clustering*: It clusters the representative vectors, and thus corresponding documents, into groups where the similarity between two vectors is defined as their cosine. The documents can be clustered into a hierarchy via top-down phases each of which uses one of the four NE-based component vectors presented above. That is, a cluster can be divided further into smaller ones in the hierarchy.

4. *Cluster representation generation*: It generates meaningful and human-understandable representations of the formed clusters, using the most significant NE feature values to label the clustered documents. In addition, the clusters are organized into a hierarchy to provide a more structural view than a flat partition of the documents.

For example, given a set of geographical documents, one can first cluster them into groups of documents about rivers and mountains, i.e., clustering with respect to entity types. Then, the documents in the river group can be clustered further into subgroups each of which is about a particular river, i.e., clustering with respect to entity identifiers. As another example of combination of clustering objectives, one can first make a group of documents about entities named "*Shenyang*", by clustering them with respect to entity names. Then, the documents within this group can be clustered further into subgroups for *Shenyang City*, *Shenyang University*, and *Shenyang Hotel*, for instance, by clustering them with respect to entity types.

In this work, we adopt the most popular partitioning technique, *k*-means ([6]), for vector clustering, but any technique could be used in the framework. We recall that, basically, the *k*-means technique keeps relocating data points into *k* clusters until the following objective function stops decreasing:

$$f = \sum_{i=1}^{k} \sum_{x_j \in c_i} |x_j - \overline{c_i}|$$

where $c_i$ is the *i*-th cluster and $\overline{c_i}$ is the average value of its data points $x_j$'s, called the centroid. In practice, for obtaining the best clustering quality, the optimal value of *k* can be determined by experiments.

Theoretically, clustering quality can be evaluated using two complementary measures: (1) *internal measure* that reflects the average semantic distance between data points within each cluster; the smaller the better; and (2) *external measure* that reflects the average semantic distance between the clusters; the larger the better. In particular, in this work, we employ the *cluster entropy* and the *class entropy* defined in [7] as the internal and external measures, respectively.

Formally, suppose a set of clusters $c_i$'s and a set of labels $l_j$'s, each of which is assigned to those data points that share the same certain feature values. Let *N* be the total number of data points, $nc_i$ be the number of data points in cluster $c_i$, $nl_j$ be the number of data points of label $l_j$, and $n_{ij}$ be the number of data points labeled $l_j$ in cluster $c_i$. Then, the cluster entropy $Ec$ and the class entropy $El$ are defined as follows:

$$Ec = -\sum_i \frac{nc_i}{N} \sum_j \frac{n_{ij}}{nc_i} \log(\frac{n_{ij}}{nc_i})$$

$$El = -\sum_j \frac{nl_j}{N} \sum_i \frac{n_{ij}}{nl_j} \log(\frac{n_{ij}}{nl_j})$$

It can be observed that, for *k*-means clustering, if the pre-specified number of clusters *k* increases, then the class entropy tends to increase but the cluster entropy tends to decrease. Meanwhile, if the value of *k* decreases, then the class entropy decreases while the cluster entropy increases. So, overall entropy can be defined as a linear combination of the cluster and class entropies as below:

$$E = \alpha.Ec + (1 - \alpha).El$$

where $\alpha$ is empirically determined. The smaller *E* is, the better clustering quality is. Ideally, all data points in each cluster have the same label, i.e., $Ec = 0$, and all data points of the same label reside in the same cluster, i.e., $El = 0$.

In our proposed NE-based clustering method, the label of a document is defined by a set of the most significant NE feature values in the document. The significance of a NE feature value is determined by its *tf.idf* weight as calculated in Section II, assumed that the higher a *tf.idf* weight is, the more significant the corresponding feature value. For example, if documents are clustered by entity types, then each document is labeled by a set of those entity types with the greatest *tf.idf* weights in that document. The label of a cluster is then constructed as the union of the sets of NE feature values that label the documents in that cluster.

In practice, a confidence threshold $T_c$ is used to filter those most significant NE feature values. That is, if the weight of a feature value in a document is greater than, or equal to, $T_c$, it dominates the meaning and should be selected as the label of the document. As for $\alpha$, the optimal value of $T_c$ is determined empirically. If all the feature value weights in a document are less than $T_c$, we take the feature value having the maximal weight as the label for the document.

IV. EXPERIMENTAL RESULTS

For experiments, our dataset consists of 960 documents collected mainly in the politics domain from BBC News. KIM ([8]) has been employed for NE recognition, whence among the collected documents 4826 entities have been recognized, covering 49 types, 4826 identifiers, and 5274 names. We have implemented the system so that users can choose different combinations of NE features for clustering in different phases. We have also tuned the values of *k* and $\alpha$ in order to get optimal clustering quality in terms of overall entropy. Our experiments have shown that clustering quality tends to reach optimum when $\alpha$ approaches 0.5, i.e., the cluster and class entropies have the same significance for the overall entropy. The experiments also point out that when $T_c$ is equal to 40% of the sum of all feature value weights in a document, the best clustering quality is obtained.

Figure 2 shows the changes of the cluster, class, and overall entropies with respect to *k*, when $\alpha$ is 0.5 and the collected documents are clustered by NE types. The optimal value of *k* in this case is 25, obtained when the overall entropy is minimal. Generally, in order to find the optimum value of *k*, we vary the value of *k* in the possible range from 1 to the total number of documents (i.e., 960 in this case), then select the value that gives the minimal overall entropy as defined in Section III.

Table I shows the optimal numbers of clusters with respect to different clustering objectives, obtained by tuning experiments. One can realize that the clustering result obtained when the chosen objective is entity type is rather good. In this case, the achieved overall entropy is less 0.3, theoretically meaning that, for each 10 documents, there are more than 7 documents clustered reasonably.

We note that the information of entity types is only available if we process the document semantics at the named entity level. Then, documents about entities that are distinct but have the same types can be grouped together. In contrast, if the documents were only processed at the keyword level, the clustering result would be like that obtained when entity occurrences were considered as keywords, i.e., each entity name as a keyword. It would be even worse, because those entities that had different types but identical name were treated the same.

Table I also shows that when clustering by entity identifiers, the best overall entropy is obtained when number of clusters is very close to that of documents, i.e. almost every document is classified into a distinct cluster. First, this reflects the fact that our randomly collected dataset is extremely inhomogeneous in terms of entity identifiers, i.e., almost each document is about a distinct entity. The clustering result, theoretically, should be improved with a dataset whose document contents are more focusing on a small set of entities. Second, clustering by entity identifiers is not the same as clustering by keywords, because one entity may appear with different names.

When clustering by entity names, the clustering result is just slightly different from that obtained by entity identifiers. It can be explained by the fact that, in the collected dataset, the entities in the documents almost have distinct names. We also note that clustering by entity names is not the same as clustering by keywords, although entity names are treated as keywords in the latter. That is because, due to the way the *tf.idf* weights are computed with respect to entity names, the aliases of each entity are taken into account, though the entity may only appear in a document under its main name. For example, when clustered by entity names, a document in which the name "*Georgia*" representing a city (in the United States) is dominant and a document in which the name "*Gruzia*" representing a country (in the former Soviet Union) is dominant could be grouped together, because "*Georgia*" is an alias of *Gruzia*.

As such, named entities introduce a new perspective to document clustering, which is complementary to keywords in satisfying user needs of organizing and viewing documents in a structured way. As explained above, a type may correspond to more than one entity identifier, an identifier to more than one name entity, and a name to more than one keyword. Therefore, in general, the homogeneity of documents with respect to entity types, identifiers, names, and keywords decreases in that order. We recall that information of named entities is not taken into account in the traditional textual information processing approaches.

Moreover, users can specify different clustering objectives as different combinations of NE features, to observe the information in their desired manners. In Table I, we present the clustering result obtained when we cluster the documents firstly by entity types and then by identifiers. In this case, the number of clusters produced in the first phase with respect to entity types is actually 3, whose numbers of sub-clusters with respect to entity identifiers are respectively 239, 320, and 306, summed up to 865. It is reasonable that the overall clustering quality is close to that of clustering by entity identifiers in one phase only. However, in terms of representation, with this two-phase clustering, we can present the clustering results in a more organized and meaningful manner to observers, as shown in Example IV.2 below.

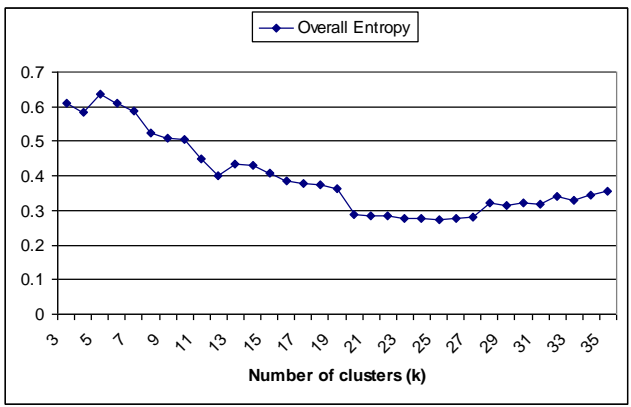

Fig. 2. Tuning the number of clusters when clustering by NE types with α = 0.5

TABLE I
OPTIMAL NUMBERS OF CLUSTERS WITH RESPECT TO DIFFERENT CLUSTERING OBJECTIVES

| Clustering Objectives | k | Cluster Entropy | Class Entropy | Overall Entropy |
|---|---|---|---|---|
| Type | 25 | 0.32 | 0.22 | 0.27 |
| Identifier | 949 | 0.00 | 0.75 | 0.37 |
| Name | 935 | 0.00 | 0.77 | 0.38 |
| Type-then-Identifier | 865 | 0.04 | 0.73 | 0.38 |

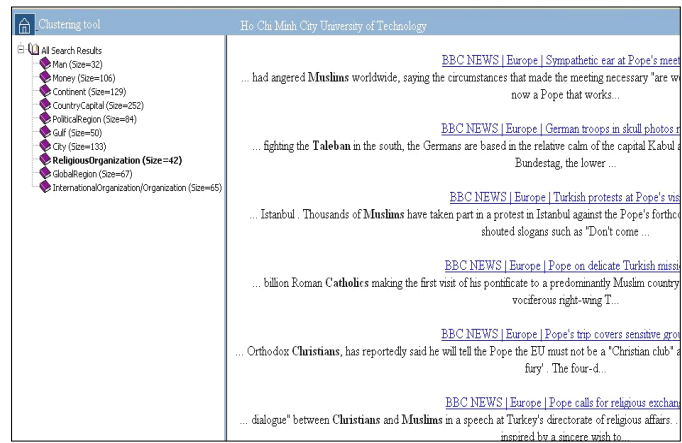

Fig. 3. A screenshot of NE-based clustering results by entity types

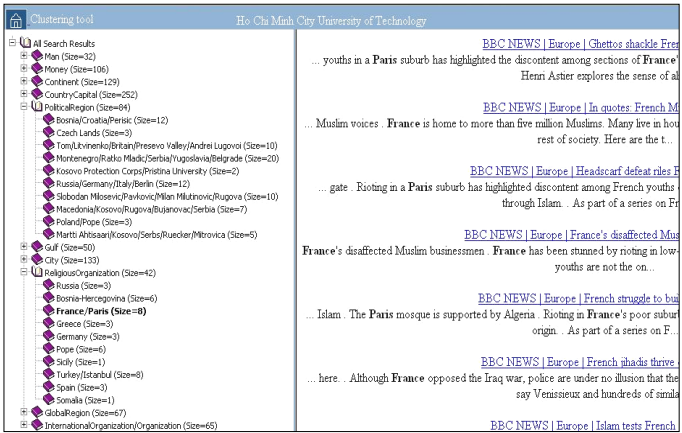

Fig. 4. A screenshot of NE-based clustering results by entity types-then-identifiers

**Example IV.1** Figure 3 shows the cluster hierarchy generated when we cluster the dataset based on entity types. As can be observed in the figure, the documents in the dataset have been classified into groups each of which is represented by a set of real world concepts such as *CountryCapital*, *PoliticalRegion*, or *ReligiousOrganization*. The label *ReligiousOrganization (Size=42)*, for instance, means that the cluster has 42 documents in which the majority of named entities are of the type *ReligousOrganization*. In fact, as shown in the right panel of the screen window, the documents in this cluster are mainly about events related to religious organizations such as *Muslim*, *Taleban*, *Catholics*, and *Christians*.

**Example IV.2** Figure 4 shows the cluster hierarchy generated when we cluster the dataset in two phases. In the first phase, the documents are clustered based on entity types, resulting in the first layer of clusters represented by a set of real world concepts, similar to those in Example IV.1. In the second phase, the documents in each generated cluster in the first phase are clustered further by entity identifiers. It generates the second layer of clusters each of which is represented by real world individuals of the corresponding concepts. Thus, clusters in the second layer help users to continue exploring information in a more concrete manner. For example, in the *ReligiousOrganization* cluster in the first layer, user can still go further to observe a more detailed classification of the documents, such as those discussing about religious conflict in *Paris*, *France*.

## V. CONCLUSION

We have presented an adaptation of the traditional keyword-based VSM with vectors over NE spaces. Each document (or query) is represented by four component vectors over the four spaces of entity names, types, name-type pairs, and identifiers, allowing searching and clustering documents by various NE features. Vector dimensional weights are computed in accordance to the *tf.idf* scheme and with respect to each of those four features of named entities. Similarity between two documents is then defined as the cosine of their representative vectors. The essential of our proposed model is that distinct features of named entities, type subsumption, and name alias are all taken into account.

We have introduced named entities into document clustering using the proposed NE-based VSM. Experiments show that NE-based clustering is complementary to keyword-based one, giving a new perspective for user needs and producing meaningful layers and groups of documents, with respect to different features of named entities mentioned in the documents. It could be applied to web-based learning where, in many subjects, named entities together with general concepts constitute the main contents of a document.

As presented in the paper, the label of a document is currently defined as the set of the dominant feature values in the document, and two labels are considered to be totally different if their two defining sets are different, though they may share most of feature values. It would be more reasonable, and bring better clustering quality, if the overlapping of those two sets is taken into account. In another aspect, combination both keyword-based clustering and NE-based clustering is to be investigated because, for instance, we may want to group documents about earthquake, by keywords, and divide the group into subgroups each of which is about earthquake in a particular country, by entity identifiers.